\documentclass[draftclsnofoot,onecolumn]{IEEEtran}
\usepackage{amsmath,amsfonts}
\usepackage{mathrsfs}
\usepackage{algorithmic}
\usepackage{algorithm}
\usepackage{array}
\usepackage{textcomp}
\usepackage{stfloats}
\usepackage{url}
\usepackage{hyperref}
\hypersetup{
	colorlinks=true,
	linkcolor=blue,
	filecolor=blue,
	urlcolor=blue,
	citecolor=blue,
}
\usepackage{xcolor}
\usepackage{verbatim}
\usepackage{graphicx}
\graphicspath{{Figures/}}
\usepackage{epstopdf}
\usepackage{cite}
\usepackage{subcaption}
\usepackage{multirow}
\usepackage{bm}
\def\BibTeX{{\rm B\kern-.05em{\sc i\kern-.025em b}\kern-.08em
		T\kern-.1667em\lower.7ex\hbox{E}\kern-.125emX}}
	
\begin{document}

\title{Hybrid Bit and Semantic Communications}
\author{\IEEEauthorblockN{Kaiwen Yu, Renhe Fan, Gang Wu, and Zhijin Qin}
\thanks{This work was supported by the Fundamental Research Funds for the Central Universities under Grant 2242022k60006.
Kaiwen Yu, Renhe Fan, and Gang Wu are with the National Key Laboratory of Wireless Communications, University of Electronic Science and Technology of China, Chengdu 611731, China. (e-mail: yukaiwen@std.uestc.edu.cn, 202321220110@std.uestc.edu.cn, wugang99@uestc.edu.cn). 
Zhijin Qin is with the Department of Electronic Engineering, Tsinghua University, Beijing 100084, China (e-mail: qinzhijin@tsinghua.edu.cn).}
}

\maketitle

\begin{abstract}
  Semantic communication technology is regarded as a method surpassing the Shannon limit of bit transmission, capable of effectively enhancing transmission efficiency. However, current approaches that directly map content to transmission symbols are challenging to deploy in practice, imposing significant limitations on the development of semantic communication. To address this challenge, we propose a hybrid bit and semantic communication system, named HybridBSC, in which encoded semantic information is inserted into bit information for transmission via conventional digital communication systems utilizing same spectrum resources. 
  The system can be easily deployed using existing communication architecture to achieve bit and semantic information transmission.
  Particularly, we design a semantic insertion and extraction scheme to implement this strategy. Furthermore, we conduct experimental validation based on the pluto-based software defined radio (SDR) platform in a real wireless channel, demonstrating that the proposed strategy can simultaneously transmit semantic and bit information.
\end{abstract}

\begin{IEEEkeywords}
Semantic communications, deep learning, bit communications, wireless image transmission, SDR platform.
\end{IEEEkeywords}

\section{Introduction}

\IEEEPARstart {A}{s} many applications generate extensive and diverse data to facilitate intelligent services, the demand for channel capacity in wireless networks is expected to experience unrestricted exponential growth. 
Conventional communication systems presently encounter bottlenecks in supporting these multimodal and voluminous data transmissions. Semantic communication technology is considered a crucial means to surpass the Shannon limit, exhibiting higher efficiency compared to traditional bit transmission systems\cite{gunduz2022beyond, lan2021semantic}. It is expected to be applicable to diverse data types and scenarios involving large data volumes\cite{qin2024computing}.

Semantic communications employ a joint source-channel coding approach for end-to-end communications\cite{bourtsoulatze2019deep}. In this context, the transmitter directly maps the source content into the symbols before the wireless channel.
The receiver then directly reconstructs these symbols distorted by wireless channel back into their original content\cite{yoo2023role}.
Consequently, the semantic encoding and decoding processes are highly dependent on the content to be transmitted.
Various modal data require semantic encoders and decoders with different architectures for effective semantic extraction and recovery\cite{xie2021deep, yu2024two, wang2022wireless}.
Traditional bit transmission methods require consistent transmission of bit sequences representing the entire source content, while semantic communications selectively transmit relevant information necessary to complete the communication task at the receiver\cite{xu2023knowledge,kurka2020deepjscc}.
This significantly reduces the size of data that needs to be transmitted, thereby significantly saving spectrum resources.

Some researches have focused on evolutionary schemes from bit communications to semantic communications. By utilizing semantic communications for transmitting large data-volume content such as augmented reality (AR), virtual reality (VR), or extended reality (XR) services, while using traditional bit communications for transmitting other small data-volume content\cite{wang2023adaptive}. This strategy leverages the high data compression feature of semantic communications to overcome difficulties associated with large data-volume transmission. Furthermore, a similar approach is to use joint source-channel coding for remote communication users, while relying on traditional bit communications for close communication users\cite{mu2022heterogeneous}. This strategy exploits the robustness of semantic communications to tackle the issue of poor channel quality for remote users. 
Overall, these strategies are equivalent to deploying additional semantic communication system outside the original bit communications network.

However, semantic communication strategy directly maps application layer data to physical layer transmission symbols, which will break the existing protocol architecture, pose a greater challenges to existing communication networks, and make it difficult to popularize applications.
3GPP prefers a gradual transition approach to technology updates. Disruptive evolution may cause huge economic losses to the industry, which is unbearable.
Given these challenges, is it possible to design a semi-evolutionary semantic communication strategy that not only improves transmission quality but also integrates well with existing communication architecture? 
A simple solution is to design a transceiver where bits and semantics coexist, and use existing digital communication architecture for wireless transmission. However, this would face some inevitable difficulties. First, how to let bit information and semantic information coexist? Secondly, how to deal with the distortion caused by source coding compression and wireless channel fading? Finally, how to recover bit and semantic information from the same spectrum resource? 
In this paper, we study a novel bit and semantic co-transmission system and the main contributions of this work are summarized as follows:

\begin{itemize}
  \item We propose a novel hybrid bit and semantic communication system, named HybridBSC, where semantic information is inserted into bit-encoded images and transmitted in traditional digital communications.
  \item We design a semantic insertion and extraction scheme so that semantic information and bit information share the same spectrum resources while countering the distortion imposed by source coding compression and wireless channel fading.
  \item We implement a HybridBSC prototype based on the pluto-based software defined radio (SDR) platform and 802.11a OFDM PHY transceiver structure \cite{ieee2000wireless} and conduct experiments in a real wireless channel. The results verify that the proposed strategy can accurately transmit semantic and bit information.
\end{itemize}

\section{The Proposed Hybrid Bit and Semantic Communication System}
\begin{figure}[t!]
	\centerline{\includegraphics[width=3.5in]{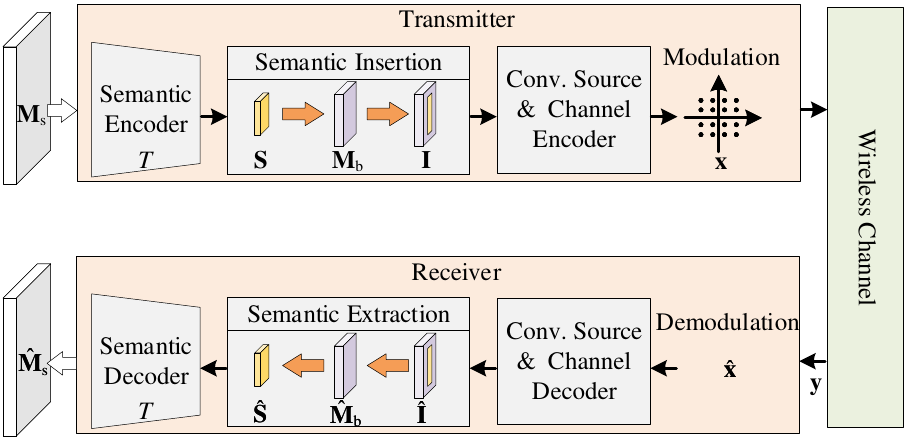}}
	\caption{The structure of the proposed HybridBSC.}
	\label{SysModel}
\end{figure}

In this section, we introduce the proposed HybridBSC. As shown in Fig.~\ref{SysModel}, we consider a semantic and bit information occupy the same spectrum resources for simultaneous transmission over the wireless channel, where the transmitter and receiver hold both conventional bit and semantic encoders and decoders, respectively. Both the semantic encoder and decoder are represented by deep neural networks.
\subsection{Transmitter}
As shown in Fig.~\ref{SysModel}, the transmitter consists of three parts: semantic encoder, conventional bit encoder, and channel encoder. We denote the semantic-encoded input images as $\mathbf M_{\rm s}$ and the bit-encoded input images as $\mathbf M_{\rm b}$. The semantic feature tensor $\mathbf S$ is extracted first as
\begin{equation}\label{se}
	\mathbf {S} = T\left( \mathbf M_{\rm s}; {\bm{\alpha }} \right),
\end{equation}
where $T\left( -; {\bm{\alpha }} \right)$ denotes the semantic encoder with learnable parameters $\bm{\alpha }$. 
The semantic encoder is trained to convert the transmitted data into encoded feature vectors, which are then ready for insertion into the image sources $\mathbf M_{\rm b}$ for conventional encoding. This insertion process can be represented as
\begin{equation}\label{be}
	\mathbf {I} = E\left( \mathbf S, \mathbf M_{\rm b}\right),
\end{equation}
where $E$ denotes the insertion function, which will be driven by the designed algorithm. This process will be described in detail later, and $\mathbf I$ denotes the hybrid information, which contains semantics and bit data. Then, after conventional source coding, channel coding and modulation, the hybrid information $\mathbf {I}$ is processed into transmitted complex symbol $\mathbf x $.

\subsection{Receiver}
The receiver also consists of three parts: semantic decoder, conventional bit decoder, and channel decoder. The received signal $\mathbf y$ can be expressed as
\begin{equation}\label{Rx}
\mathbf {y} = {\mathbf h} {\mathbf x} + \mathbf{n},
\end{equation}
where $\mathbf h$ is the coefficients of a linear channel. If a additive white Gaussian noise (AWGN) channel, the channel coefficients equals $1$. If a Rayleigh fading channel, the channel coefficients follows $\mathcal{N}\left(0,1\right)$.
$\mathbf n \in \mathcal N\left( {0,{\sigma ^2}} \right)$ is the additive circular complex Gaussian noise with zero mean and variance $\sigma^2$. The recovered signals are
\begin{equation}\label{Rx}
\hat {\mathbf x} = {\left( {{ {\mathbf h}^H}{\mathbf h}} \right)^{ - 1}}{{\mathbf h}^H} \mathbf y = \mathbf x + \hat {\mathbf n},
\end{equation}
where $\hat {\mathbf x}$ is the estimated information, $\hat {\mathbf n}={\left( {{ {\mathbf h}^H}{\mathbf h}} \right)^{ - 1}}{{\mathbf h}^H} \mathbf n$ is the impact of noise. Then, after demodulation, conventional channel decoding and source decoding, the hybrid information $\hat{\mathbf I}$ is recovered from the $\hat {\mathbf x}$. By proceeding the semantic extraction process, the receiver can reconstruct the semantic feature tensor $\hat{\mathbf S }$. Finally, the semantic decoder generates the reconstructed output $\hat{\mathbf M}_{\rm s}=R\left(\hat{\mathbf S };\bm \gamma\right)$, where $R\left(-;\bm \gamma\right)$ denotes the semantic decoder with learnable parameters $\bm \gamma$.

\section{The Proposed Semantic Insertion and Extraction Algorithms}
In the considered framework, semantic information and traditional bit information share the same time and frequency resources for transmission. To achieve this goal, we insert semantic information into the traditional bit-encoded information to generate hybrid data to be transmitted. However, these data will be distorted by the source encoding and wireless channel fading, which poses a great challenge to the stability of the hybrid data. To solve this problem, we propose a semantic insertion and extraction algorithm that can resist the effects of source compression and wireless channel fading, and accurately separate the semantic information from the bit data at receiver, thus realizing efficient and reliable semantic and bit transmission.

\subsection{Semantic Insertion Process}
To ensure the stability of semantic insertion, we use color space transformation to convert the source image $\mathbf M_{\rm b}$ that requires bit transmission, from RGB to YCBCR space. Based on ITU.BT-601 specification, the conversion process is as follows:
\begin{equation}\label{ColorTrans}
	\left[ \begin{array}{ccc}
		\!\!\!	\mathbf Y \!\!\! \\ 
		\!\!\!	{Cr}  \!\!\!    \\ 
		\!\!\!	{Cb}  \!\!\!
	\end{array} \right] \!=\! \left[ {\begin{array}{ccc}
		\!\!\!	{0.257}&{0.504}&{0.098}   \!\!\!\\ 
		\!\!\!	{-0.148}&{-0.291}&{0.439} \!\!\!\\ 
		\!\!\!	{0.439}&{-0.368}&{-0.071} \!\!\!
	\end{array}} \right]\left[ {\begin{array}{ccc}
		\!\!\!	R \!\!\!\\
		\!\!\!	G \!\!\!\\ 
		\!\!\!	B \!\!\!
	\end{array}} \right]\! +\! \left[ {\begin{array}{ccc}
		\!\!\!	{16}  \!\!\!\\ 
		\!\!\!	{128} \!\!\! \\ 
		\!\!\!	{128} \!\!\!
	\end{array}} \right],
\end{equation}
where $\mathbf Y$ channel represents the original image luminance, $Cb$ channel represents the image blue chroma, and $Cr$ channel represents the image red chroma. 

The bit source image $\mathbf M_{\rm b}$ is decomposed into four frequency bands after 2-dimensional discrete wavelet transform (DWT), $\rm FW()$: low-frequency part (LF), horizontal high-frequency part, vertical high-frequency part, and diagonal high-frequency part. The low-frequency part contains the main content of the image and is very similar to the original image, while the high-frequency part can be considered as redundant noise in the image.
We divide the LF part into $4 \times 4$ blocks and apply a discrete cosine transform (DCT), denoted as $\rm FC()$, to the each block, followed by a singular value decomposition (SVD) on the transformed output. The whole process can be expressed as
\begin{equation}\label{BDP}
	[\mathbf{U}_{\rm b},\mathbf{Z}_{\rm b},\mathbf{V}_{\rm b}] =\mathrm{SVD} \left( {\mathrm{FC}_{4 \times 4}\left( {\mathrm{FW}{{\left( \mathbf Y \right)}_{\rm LF}}} \right)} \right),
\end{equation}
where the singular value matrix $\mathbf Z_{\rm b}$ is a diagonal matrix, $\mathbf U_{\rm b}$ and $\mathbf V_{\rm b}$ are the orthogonal unitary matrices. Then, we perform the following operations on ${\mathbf Z_{\rm b}}$ to obtain the new $\tilde{\mathbf Z}_{\rm b}$:
\begin{equation}\label{Insertion}
{\left\| \tilde{\mathbf{Z}}_{i,j}^{\rm b} \right\|_{2'}} = \left\{ {\begin{array}{*{20}{c}}
			{\alpha \eta } \\ 
			{\alpha \left( {\eta  + 1} \right)} \\ 
			{\alpha \left( {\eta  + 1} \right)} \\ 
			{\alpha \eta } 
		\end{array}\begin{array}{*{20}{c}}
			{{\text{if }}\mathbf S_{i,j}^q= 0,\eta \; \text{mod} \; 2 == 1,} \\ 
			{{\text{if }}\mathbf S_{i,j}^q= 0,\eta \; \text{mod} \; 2 == 0,} \\ 
			{{\text{if }}\mathbf S_{i,j}^q= 1,\eta \; \text{mod} \; 2 == 1,} \\ 
			{{\text{if }}\mathbf S_{i,j}^q= 1,\eta \; \text{mod} \; 2 == 0,} 
	\end{array}} \right.
\end{equation}
where $\alpha$ is the insertion factor, $\eta  = \left\lceil {{{\left\| \mathbf{Z}_{i,j}^{\rm b} \right\|_{2'}} \mathord{\left/{\vphantom {X {}}} \right.\kern-\nulldelimiterspace} {\alpha}}} \right\rceil $. $\tilde{\mathbf{Z}}_{i,j}^{\rm b}$ denotes the $(i,j)$-th element of singular value matrix $\tilde{\mathbf{Z}}_{\rm b}$. ${\left\| \mathbf \Gamma \right\|_{2'}} = {\text{su}}{{\text{p}}_{x \in {R^n \backslash 0}}}\left( {{{\left| {\mathbf \Gamma x} \right|} \mathord{\left/{\vphantom {{\left| {\Gamma x} \right|} x}} \right.\kern-\nulldelimiterspace} \left|x\right|}} \right)$ denotes the largest singular value of $\mathbf \Gamma$, $\mathbf S_{i,j}^q = Q\left( \mathbf S_{i,j} \right)$ is the $q$-bit quantization output. Without loss of generality, we use $q = 4$ bit for quantization. 

Next, we execute inverse SVD, DCT and DWT to complete the semantic insertion process as shown as follows:
\begin{equation}\label{IDCTIDWT}
	\tilde {\mathbf Y} =  {\mathrm{IFW}\left( {\mathrm{IFC}_{4 \times 4}{{\left( \mathbf{U}_{\rm b}\tilde {\mathbf {Z}}_{\rm b}\mathbf{V}_{\rm b}^T\right)}}} \right)},
\end{equation}
where $\mathrm{IFW}()$ denotes the inverse DWT transform, $\mathrm{IFC}_{4 \times 4}()$ means performing an inverse DCT transform on each ${4 \times 4}$ block.
Then, we transform $\tilde {\mathbf Y}$ back to RGB space to get the hybrid data $ {\mathbf{I}}$.

\subsection{Semantic Extraction Process}
After the receiver recovers the hybrid data $\hat{\mathbf I}$, it needs to perform semantic extraction process to separate the semantic information from the $\hat{\mathbf I}$.

Similarly, the hybrid data $\hat{\mathbf I}$ is color space transformed to obtain $\hat{\mathbf Y}$, which is then subjected to a 2-dimensional DWT. The resulting LF part is partitioned into $4 \times 4$ blocks, and a DCT is applied to each block, followed by SVD. The whole process can be expressed as
\begin{equation}\label{SEP}
	[\hat {\mathbf U}_{\rm b},\hat {\mathbf {Z}}_{\rm b},\hat {\mathbf V}_{\rm b}] =\mathrm{SVD} \left( {\mathrm{FC}_{4 \times 4}\left( {\mathrm{FW}{{\left( \hat{\mathbf Y} \right)}_{\rm LF}}} \right)} \right),
\end{equation}
where $\hat {\mathbf {Z}}_{\rm b}$ is the recovered singular value matrix, $\hat {\mathbf U}_{\rm b}$ and $\hat {\mathbf V}_{\rm b}$ are the recovered orthogonal unitary matrices. 
And semantic feature extraction is performed according to the following formula:
\begin{equation}\label{SEP}
\hat{\mathbf {{S}}}_{i,j}^q = \left\{ {\begin{array}{*{20}{c}}
		1 \\ 
		0 
\end{array}} \right.\begin{array}{*{20}{c}}
	{{\text{if }} \hat \eta \; \text{mod} \; 2 == 0,} \\ 
	{{\text{if }} \hat \eta \; \text{mod} \; 2 == 1,} 
\end{array}
\end{equation}
where $\hat \eta = \left\lceil {{{\left\| \hat { \mathbf {Z}}_{i,j}^{\rm b} \right\|_{2'}} \mathord{\left/{\vphantom {X {}}} \right.\kern-\nulldelimiterspace} {\alpha}}} \right\rceil$, $\hat{\mathbf {{S}}}_{i,j}^q$ is the $(i,j)$-th element of quantized semantic information $\hat{ \mathbf {{S}}}^q$.
Then, we obtain the final semantic information $\hat{\mathbf S}$ by dequantization process $\hat{\mathbf S} = Q^{-1}\left(\hat{{\mathbf {{S}}}}^q\right)$.
It is worth noting that the insertion factor $\alpha$ is known to the transceiver, so the receiver can accurately extract semantic information from the hybrid data $\hat{\mathbf I}$.
The proposed hybridBSC is summarized in Algorithm.~\ref{HBSC}.

\begin{algorithm}[htbp]
\caption{The proposed HybridBSC system.}
\begin{algorithmic}[1]
\label{HBSC}
\STATE \textbf{Input:}Bit and semantic-transmitted images $\mathbf M_{\rm b}$ and $\mathbf M_{\rm s}$.
\STATE \textbf{Output:} The recovered images  $\hat {\mathbf M}_{\rm b}$ and $\hat {\mathbf M}_{\rm s}$;

\STATE Semantic encoding: $\mathbf {S} = T\left( \mathbf M_{\rm s}; {\bm{\alpha }} \right)$,
\STATE Semantic insertion: $\mathbf {I} = E\left( \mathbf S,\mathbf M_{\rm b}\right)$,
\STATE Traditional source encoding and channel encoding,
\STATE Though wireless channel: $\mathbf {y} = {\mathbf h} {\mathbf x} + \mathbf{n}$,
\STATE Traditional channel decoding and source decoding,
\STATE Semantic extraction:  $\hat {\mathbf {S}} = E^{-1}\left( \hat {\mathbf I}\right)$,
\STATE Semantic decoding: $\hat{\mathbf M}=R\left(\hat{\mathbf S };\bm \gamma\right)$

\end{algorithmic}
\end{algorithm}

%


\section{Demo and Simulations}
In this part, we evaluate the proposed HybridBSC in terms of peak signal-to-noise ratio (PSNR) and structure similarity (SSIM) performance over different channels. We use the MNIST handwritten digit dataset for semantic transmission and the USC-SIPI image dataset \cite{weber2006usc} for bit transmission. In the experiment, the semantic encoding consists of four 2-dimensional convolutional layers (Conv2D) with 32, 16, 16, and 1 neurons. The Conv2Ds are with a $3 \times 3$ kernels and followed by a ReLU activation function.
The semantic decoding also consists of four Conv2Ds with 32, 16, 16, and 1 neurons. The first three Conv2Ds are with a $3 \times 3$ kernels and followed by a ReLU activation function, and the last Conv2D is with a $3 \times 3$ kernels and followed by a sigmoid activation function. The weights of models are updated by the Adam optimizer. We set the training batch size to 256.
The mean square error (MSE) as a loss function. The learning rate is $\delta=1e^{-4}$. The epoch for training is 20. The insertion factor $\alpha$ is 14.
It is worth noting that we only focus on the initial proof-of-concept of the proposed HybridBSC, and thus the specific neural network can be adjusted.
We select joint photographic experts group (JPEG) as the conventional source coding and low-density parity-check (LDPC) as the conventional channel coding, and name it ``JPEG-LDPC" as the baseline strategy.

\vspace{-5mm}
\subsection{Numerical Results}
\begin{figure}[htbp]
	\centerline{\includegraphics[width=2.9in]{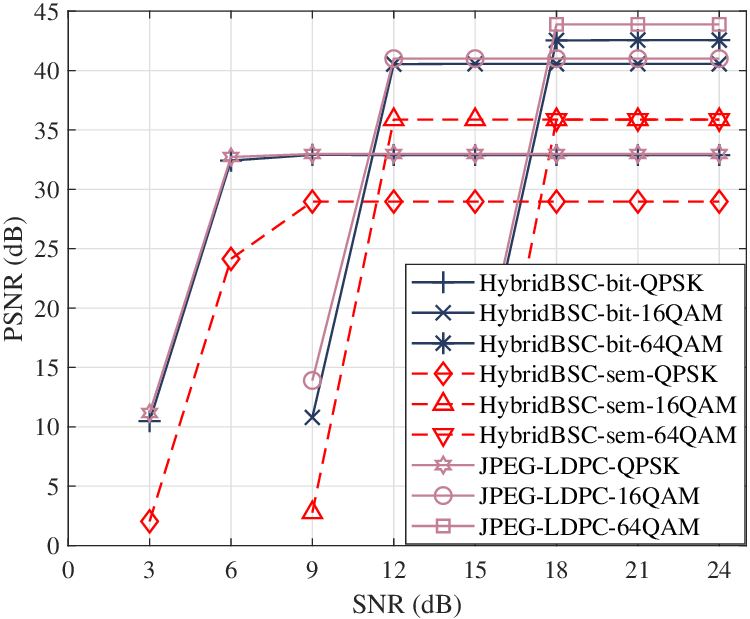}}
	\caption{The PSNR score versus SNR in AWGN channel.}
	\label{HBSC-PSNR-AWGN}
\end{figure}

\begin{figure}[htbp]
	\centerline{\includegraphics[width=2.9in]{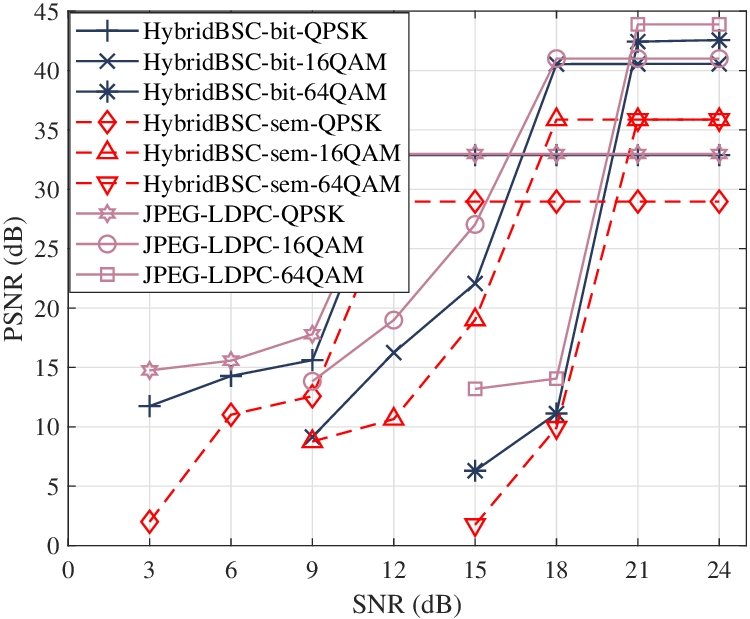}}
	\caption{The PSNR score versus SNR in Rayleigh channel.}
	\label{HBSC-PSNR-Ray}
\end{figure}

Fig.~\ref{HBSC-PSNR-AWGN} and Fig.~\ref{HBSC-PSNR-Ray} illustrate the PSNR performance comparison for various modulation methods under the same number of transmission symbols on AWGN and Rayleigh channels, respectively. 
The ``HybridBSC-bit'' denotes the performance of bit-transmitted images, and ``HybridBSC-sem'' denotes the performance of semantic-transmitted images.
These results demonstrate the efficacy of our HybridBSC in accurately transmitting both types of information.
Moreover, the bit transmission performance of the proposed HybridBSC is very close to that of the JPEG-LDPC under three modulation cases, which proves that the proposed semantic insertion and extraction strategy hardly impairs the bit transmission performance.
In addition, the semantic transmission performance is slightly lower than the bit transmission performance in different modulation cases. Several factors contribute to this disparity. Firstly, semantic information is inserted in the image intended for bit transmitted, necessitating accurate recovery of the bit image prior to semantic extraction. Secondly, the performance gap between semantic and bit information is primarily influenced by the insertion factor in the semantic insertion algorithm, which can be adjusted to balance their performance. Lastly, the semantic-transmitted performance is affected by the semantic encoding and decoding structure. While more complex structures may offer superior performance, this aspect falls outside the scope of this article, which primarily emphasizes the simultaneous and precise transmission of semantic and bit information.

\begin{figure}[htbp]
	\centerline{\includegraphics[width=2.9in]{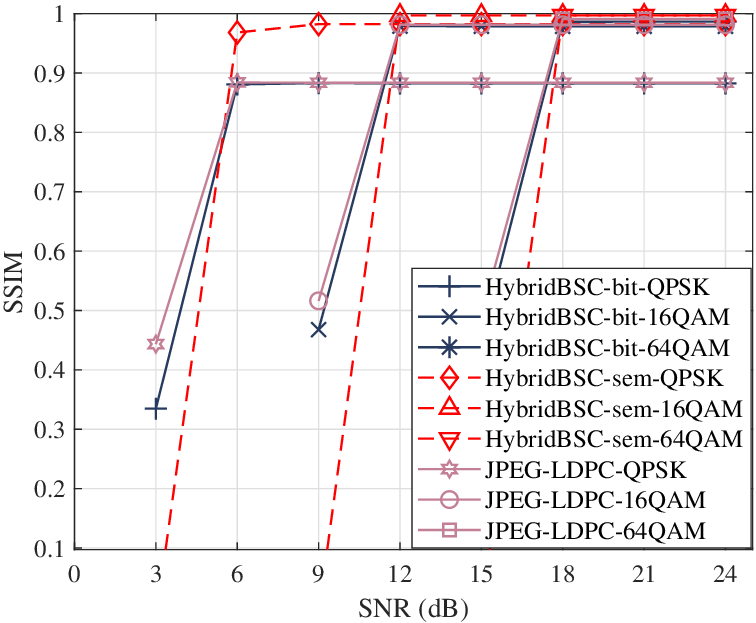}}  
	\caption{The SSIM score versus SNR in AWGN channel.}
	\label{HBSC-SSIM-AWGN}
\end{figure}

\begin{figure}[htbp]
	\centerline{\includegraphics[width=2.9in]{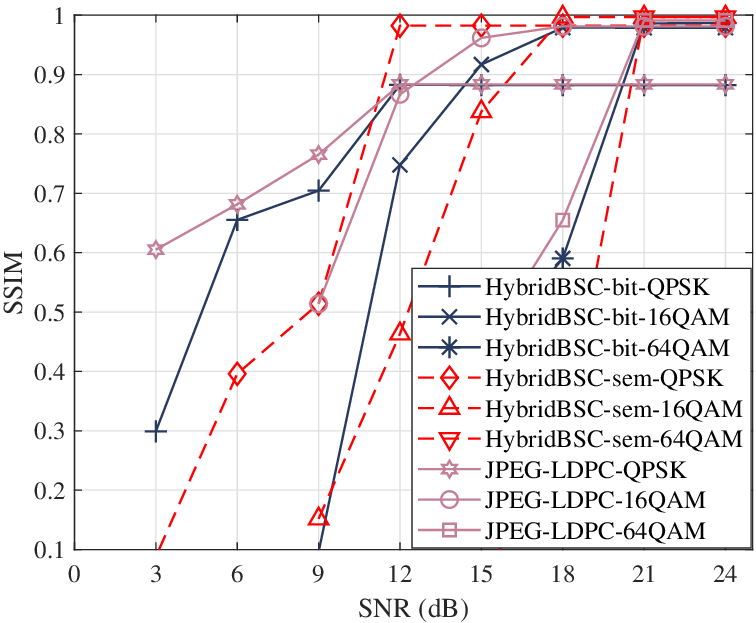}}
	\caption{The SSIM score versus SNR in Rayleigh channel.}
	\label{HBSC-SSIM-Ray}
\end{figure}

Fig.~\ref{HBSC-SSIM-AWGN} and Fig.~\ref{HBSC-SSIM-Ray} depict the SSIM performance comparison for various modulation methods under the same number of transmission symbols on AWGN and Rayleigh channels, respectively.
From the figure, the bit transmission performance with QPSK exhibits lower efficiency compared to semantic transmission performance, attributed to JPEG compression damaging its structure.
However, both bit and semantic information can be transmitted accurately across all modulation cases. In the Rayleigh channel, particularly at the low SNR regimes, both bit and semantic information transmission performance are lower than those under the AWGN channel. Nevertheless, at high SNR regimes, the best SSIM performance is achieved in all modulation cases. Notably, the SSIM performance of the proposed HybridBSC is very close to that of conventional digital transmission.
Furthermore, at the high SNR regimes, the semantic-transmitted performance with 16QAM and 64QAM is equivalent, as the insertion factor can compensate for certain JPEG compression impairments.

\subsection{Experience Results}
We implement a wireless HybridBSC prototype based on the ADALM pluto SDR platform and 802.11a OFDM PHY \cite{ieee2000wireless} transceiver architecture for experiments in a real wireless channel. 
According to IEEE 802.11a, the transmitter sequentially converts binary bitstreams into MAC Service Data Units (MSDU), MAC Protocol Data Units (MPDU), and Physical Service Data Units (PSDU), followed by the completion of MAC layer procedures. Subsequently, the transmitter conducts physical operations on the data to produce a Physical Protocol Data Unit (PPDU) for transmission. Similarly, the receiver performs the reverse process of physical layer and MAC to recover the transmitted data.
The experiments were performed in line-of-sight (LoS) environments. The transmission was carried out with a bandwidth of 20 MHz and a center frequency of 2.48 GHz. The gains of transmitter and receiver are both 0 dB.

\begin{figure*}[htbp]
	\centering
	\subcaptionbox{\label{ExperimentalScenario}}[.32\linewidth]
		{\includegraphics[width=0.3\textwidth]{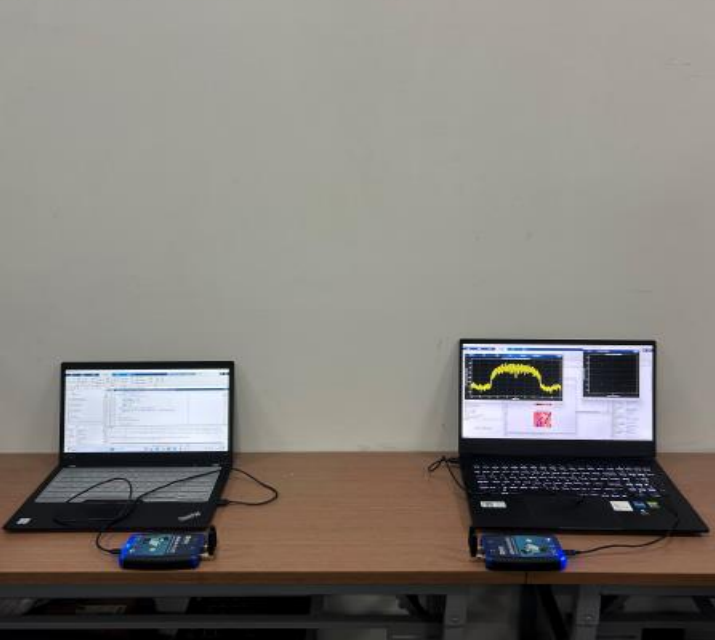}}
	\subcaptionbox{\label{ConstellationDiagram}}[.32\linewidth]
		{\includegraphics[width=0.3\textwidth]{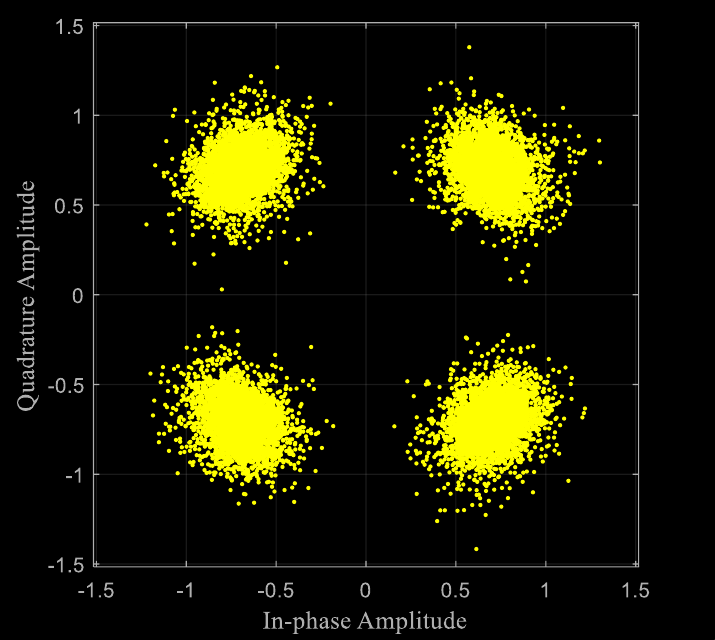}}
	\subcaptionbox{\label{SpectrumDiagram}}[.32\linewidth]
		{\includegraphics[width=0.3\textwidth]{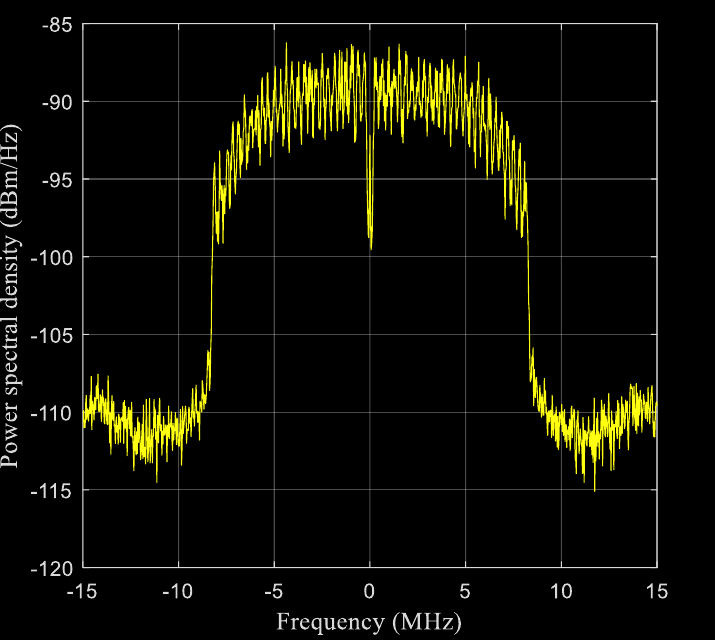}}
	
	\caption{The pluto-based SDR wireless HybridBSC prototype. (a) Real experimental scenario of the proposed HybridBSC in a indoor LoS environment; (b) Example of I/Q signals transmitted with QPSK modulation. (c) Example of received baseband WLAN signal spectrum in the wireless testbed.}\label{ExperienceScen} 
\end{figure*}

\begin{figure}[t!]
	\centering
	\subcaptionbox{The transmitted data (Top: bit, bottom: semantic)\label{OriginalImage}}
	[1\linewidth]
	{
		{\includegraphics[width=0.1125\textwidth]{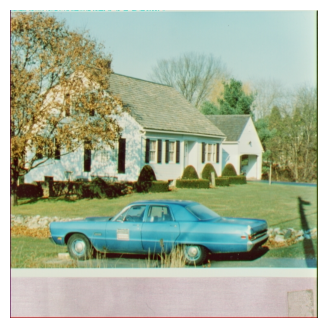}
		\includegraphics[width=0.1125\textwidth]{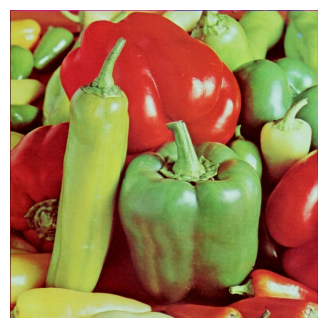}
		\includegraphics[width=0.1125\textwidth]{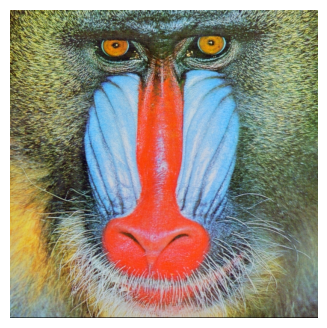}	
		\includegraphics[width=0.1125\textwidth]{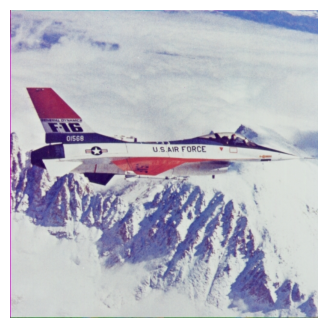}} 
		
	    {\includegraphics[width=0.1125\textwidth]{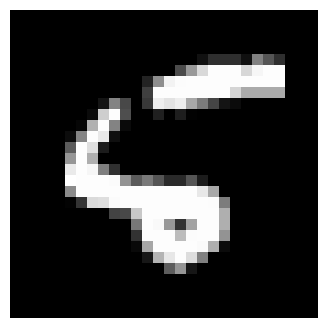}
		\includegraphics[width=0.1125\textwidth]{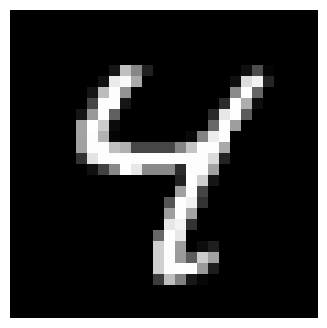}
		\includegraphics[width=0.1125\textwidth]{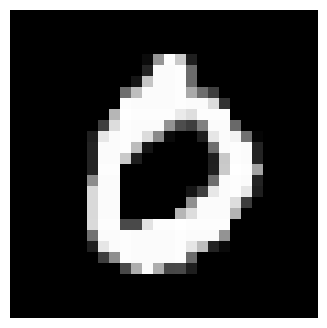}	
		\includegraphics[width=0.1125\textwidth]{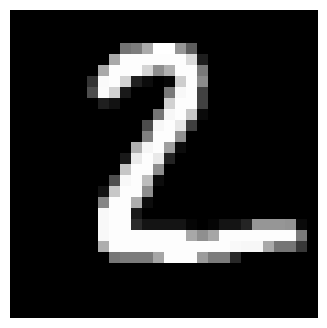}}
	}
	
	\centering
	\subcaptionbox{The recovered data (Top: bit, bottom: semantic)\label{RecoverImage}}
	[1\linewidth]
	{
		{\includegraphics[width=0.1125\textwidth]{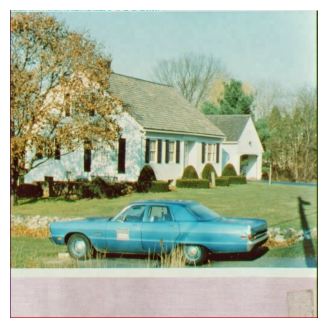}
			\includegraphics[width=0.1125\textwidth]{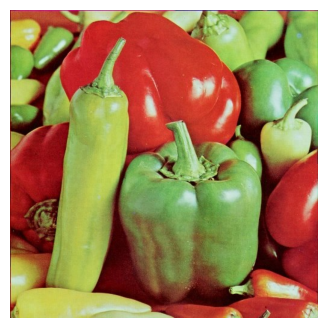}
			\includegraphics[width=0.1125\textwidth]{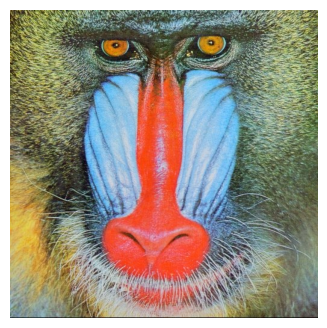}	
			\includegraphics[width=0.1125\textwidth]{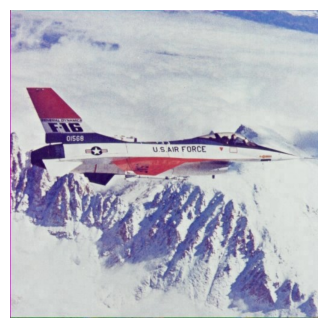}} 
		
		{\includegraphics[width=0.1125\textwidth]{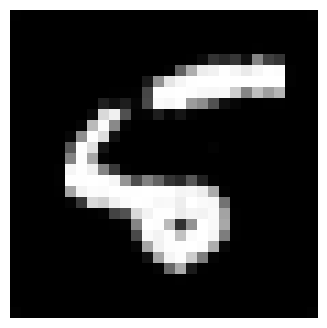}
			\includegraphics[width=0.1125\textwidth]{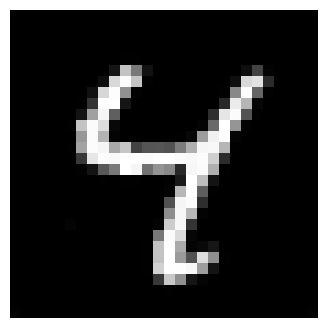}
			\includegraphics[width=0.1125\textwidth]{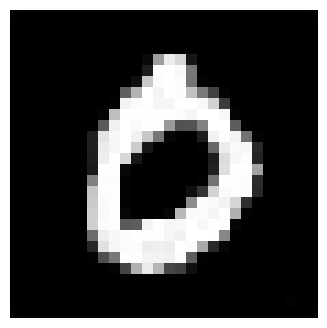}	
			\includegraphics[width=0.1125\textwidth]{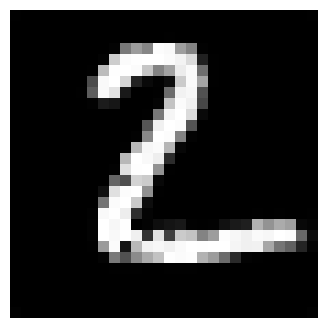}}
	}
	
	\caption{Examples of transmitted and recovered images in a real wireless channel using the designed HybridBSC prototype.}
	\label{ExperienceResult}
\end{figure}

Fig.~\ref{ExperienceScen} shows the experience scenario of the pluto-based SDR HybridBSC prototype. Specifically, Fig.~\ref{ExperienceScen}(\subref{ExperimentalScenario}) shows the real-time experience environment of the designed HybridBSC prototype, Fig.~\ref{ExperienceScen}(\subref{ConstellationDiagram}) shows the constellation diagram of QPSK modulation, Fig.~\ref{ExperienceScen}(\subref{SpectrumDiagram}) shows the received baseband wireless local area network (WLAN) signal spectrum. We first encode the images to be transmitted using a predefined encoder of the host PC, and then transmit the encoded images over a wireless channel via pluto to another PC, where the signal is decoded to produce the output images.
The output images are illustrated in Fig.~\ref{ExperienceResult}, with Fig.~\ref{ExperienceResult}(\subref{OriginalImage}) displaying the transmitted images and Fig.~\ref{ExperienceResult}(\subref{RecoverImage}) showcasing the recovered images. In these figures, the top row corresponds to the bit-transmitted images, while the bottom row represents the semantic-transmitted images.
Specifically, the average PSNR and SSIM performance for both transmitted and recovered bit images is 30.61 dB and 0.86, respectively, while for semantic images, the average PSNR is 31.30 dB and average SSIM is 0.99. This demonstrates that the proposed semantic insertion and extraction strategies effectively mitigate damage from the source encoding and real wireless channels. Additionally, it verifies the capability of the designed prototype to simultaneously transmit both bit and semantic information.

\vspace{-2mm}
\section{Conclusion}
In this article, we have studied the HybridBSC system and proposed a semantic insertion and extraction scheme for the semantic information.
We verified the system's availability in real-world wireless channels by conducting extensive experiments on a pluto-based wireless HybridBSC testbed.
The simulation and experience results show that the proposed HybridBSC is capable of transmitting bit information and semantic information simultaneously using conventional transceiver schemes in the same time-frequency resources. We hope our work provides some insights as well as facilitates the development of semantic communications.


\bibliography{Refer_HBSC}
\bibliographystyle{IEEEtran}

\end{document}